%% file: main.tex
\documentclass[conference]{IEEEtran}

\usepackage{cite}
\usepackage{amsmath,amssymb,amsfonts}
\usepackage{algorithmic}
\usepackage{graphicx}
\usepackage{textcomp}
\usepackage{comment}
\usepackage[nolist]{acronym}
\usepackage[nolist]{acronym}
\usepackage[normalem]{ulem}
\usepackage{comment}
\usepackage{color,soul}
\usepackage{multirow}
\usepackage{multicol,multirow,colortbl}
\usepackage[caption=false,font=footnotesize,labelfont=sf,textfont=sf]{subfig}
\usepackage{algorithm}
\usepackage[dvipsnames]{xcolor}
\usepackage{lmodern,textcomp}
\usepackage{adjustbox}

\UseRawInputEncoding

\input{acronym}

\begin{document}
\bstctlcite{IEEEexample:BSTcontrol}
%Cooperative Driving Strategies for CAVs at Signal-Free Intersections
% TITOLO ALTERNATIVO: 
\title{Hardware-in-the-Loop Demo of a Cooperative Intersection where Cars don't Stop}

%\author{Authors}
%\begin{comment}
\author{\IEEEauthorblockN{
Lorenzo Farina\IEEEauthorrefmark{1}\IEEEauthorrefmark{2},
Matteo Piccoli\IEEEauthorrefmark{1}, 
Salvatore Iandolo\IEEEauthorrefmark{3},
Antonio Solida\IEEEauthorrefmark{3},
Carlo Augusto Grazia\IEEEauthorrefmark{3},\\
Marco Rapelli\IEEEauthorrefmark{4},
Francesco Raviglione\IEEEauthorrefmark{4},
Claudio Casetti\IEEEauthorrefmark{5},
and Alessandro Bazzi\IEEEauthorrefmark{1}\IEEEauthorrefmark{2}
}\\
\IEEEauthorblockA{\IEEEauthorrefmark{1}DEI, Universit\`a di Bologna, 40136 Bologna, Italy}
\IEEEauthorblockA{\IEEEauthorrefmark{2}National Laboratory of Wireless Communications (WiLab), CNIT, 40136 Bologna, Italy}
\IEEEauthorblockA{\IEEEauthorrefmark{3}DIEF, University of Modena and Reggio Emilia, 41125 Modena, Italy}
\IEEEauthorblockA{\IEEEauthorrefmark{4}Dipartimento di Elettronica e Telecomunicazioni, Politecnico di Torino, 10129 Torino, Italy}
\IEEEauthorblockA{\IEEEauthorrefmark{5}Dipartimento di Automatica e Informatica, Politecnico di Torino, 10129 Torino, Italy}
%\IEEEauthorblockA{\IEEEauthorrefmark{3}Department of Control and Computer Engineering (DAUIN), Politecnico di Torino, 10129 Torino, Italy}
%\IEEEauthorblockA{\IEEEauthorrefmark{4} National Research Council (CNR) - IEIIT, 40136 Bologna, Italy}
%\thanks{Corresponding author: (email: ).}
}
%\end{comment}

%\address[1]{CNR-IEIIT and CNIT/WiLab, Bologna, Italy; (e-mail: vittorio.todisco@ieiit.cnr.it, stefania.bartoletti@ieiit.cnr.it)} 
%\address[2]{University Mediterranea of Reggio Calabria and CNIT, Reggio Calabria, Italy; (e-mail: claudia.campolo@unirc.it, antonella.molinaro@unirc.it )}
%\address[3]{Centrale Sup\'elec/L2S, Universit\'e Paris-Saclay, Gif-sur-Yvette,  France; (e-mail: antoine.berthet@supelec.fr)}
%\address[4]{Universit\`a di Bologna and CNIT/WiLab, Bologna, Italy;  (e-mail:  alessandro.bazzi@unibo.it)}

\markboth{Journal of \LaTeX\ Class Files,~Vol.~14, No.~8, August~2015}%
{Shell \MakeLowercase{\textit{et al.}}: Bare Demo of IEEEtran.cls for IEEE Transactions on Magnetics Journals}



% for Transactions on Magnetics papers, we must declare the abstract and
% index terms PRIOR to the title within the \IEEEtitleabstractindextext
% IEEEtran command as these need to go into the title area created by

% \input{Acronyms.tex}
% \input{definitions.tex}

\IEEEtitleabstractindextext{%
\begin{abstract} 
%The sector of road transportation is going through a radical change, with a future where all vehicles will be autonomous and connected. A crucial point of this future scenario is represented by the intersections, where instead of traffic lights and rights of way, algorithms and protocols will be implemented to optimize the traffic flows. In this context, we are developing a testbed with mini-car 1:10 equipped with wireless connectivity and a controller at the side of the road, to demonstrate the feasibility of intersections where the cars are coordinated and do not need to stop. This demo, in particular, will show the \ac{HiL} implementation of these intelligent intersections, using the traffic simulator SUMO to reproduce the position of the vehicles. Real hardware and software designed to be mounted on vehicles and other road devices will be exploited to reproduce in real-time the exchange of information and implement the cooperative algorithm. The vehicles, moving in different scenarios and going through different paths, exchange information with the controller and optimize their speed to avoid stopping.
The road transportation sector is undergoing a radical transformation, working for a future where all vehicles will be autonomous and connected. A key challenge in this vision concerns intersections, which will no longer rely on traffic lights or traditional right-of-way rules. Instead, algorithms and communication protocols will be employed to optimize traffic flow. In this context, we are developing a testbed where cars equipped with wireless connectivity exchange information with a remote controller and coordinate their maneuvers. As the vehicles navigate through different types of intersections with various routes, they communicate with the controller, adjusting their speed to pass through intersections without stopping. This demo specifically showcases a hardware-in-the-loop (HiL) implementation of such intelligent intersection, using the SUMO traffic simulator to emulate vehicle positions. Real hardware and software components, which are designed for use in vehicles and roadside infrastructure, are employed to enable real-time data exchange of contextual and control information, running the cooperative coordination algorithm. 
\end{abstract}
%\begin{IEEEkeywords}
%Vehicle-to-everything; Cooperative-Intelligent Transport System; Low-cost hardware; Open-source software. %\hl{non necessario in conf paper}
%\end{IEEEkeywords}
}

\maketitle

\IEEEdisplaynontitleabstractindextext

% \fi
\IEEEpeerreviewmaketitle

\acresetall

%\hl{NOTE: scadenza 20 luglio, max 3 pagine}

\section{Introduction and motivations}

The first fully autonomous vehicles are already on the road and increasing effort is being spent to make them also connected to exchange information. With connectivity, vehicles will coordinate their maneuvers and optimize the road use guaranteeing safety and passengers' comfort. In this context, one of the most critical scenarios is the intersection. In the future, the movements at the intersection will be coordinated in order to optimize the crossing, overall reducing energy consumption and travel time. 

% \begin{figure}[t]
% \centering
% \includegraphics[width=80mm,draft=true]{Figures/fig_F1.eps}
% \caption{PoC \hl{figura/e da aggiungere. foto? schema?}} %\hl{decidiamo se vogliamo evidenziare i box con i componenti, intanto allego la versione raw}}
% \label{fig:PoC}
% \end{figure}

In the Italian national project Restart MoVeOver, we are implementing algorithms that allow vehicles to cross the intersections without stopping. While the topic is being studied since a few years from a theoretical point of view, we are now moving to real implementations where protocols are developed and real hardware is used for communication. 

In this demo, we show the \ac{HiL} implementation of smart intersections where vehicles do not need to stop. The position of the vehicles in the scenario is simulated using the traffic simulator SUMO \cite{SUMO2018}, but everything else is hardware and software that can be mounted on real cars. In particular, the car \ac{OBU} is implemented using a Raspberry Pi with the \ac{V2X} standard compliant software OScar \cite{OScar_paper_2024}, which is shown to allow communications up to approximately 500~m \cite{farina2025lowcostcitsstations}. The \ac{OBU} exchanges information with a controller that is reached through a \ac{RSU}, also implemented using a Raspberry Pi with the same software. The controller implements the algorithm detailed in \cite{10575944}, which optimizes the speed of each approaching vehicle so that they do not need to stop. This solution is shown to significantly reduce both travel times and emissions with respect to traditional intersection management methods, such as using traffic lights or rights of way. %The overall demo is shown in Fig.~\ref{fig:PoC}.

\section{Demo description}

\begin{figure}[t]
\centering
\subfloat[]{\includegraphics[width=\columnwidth,draft=false]{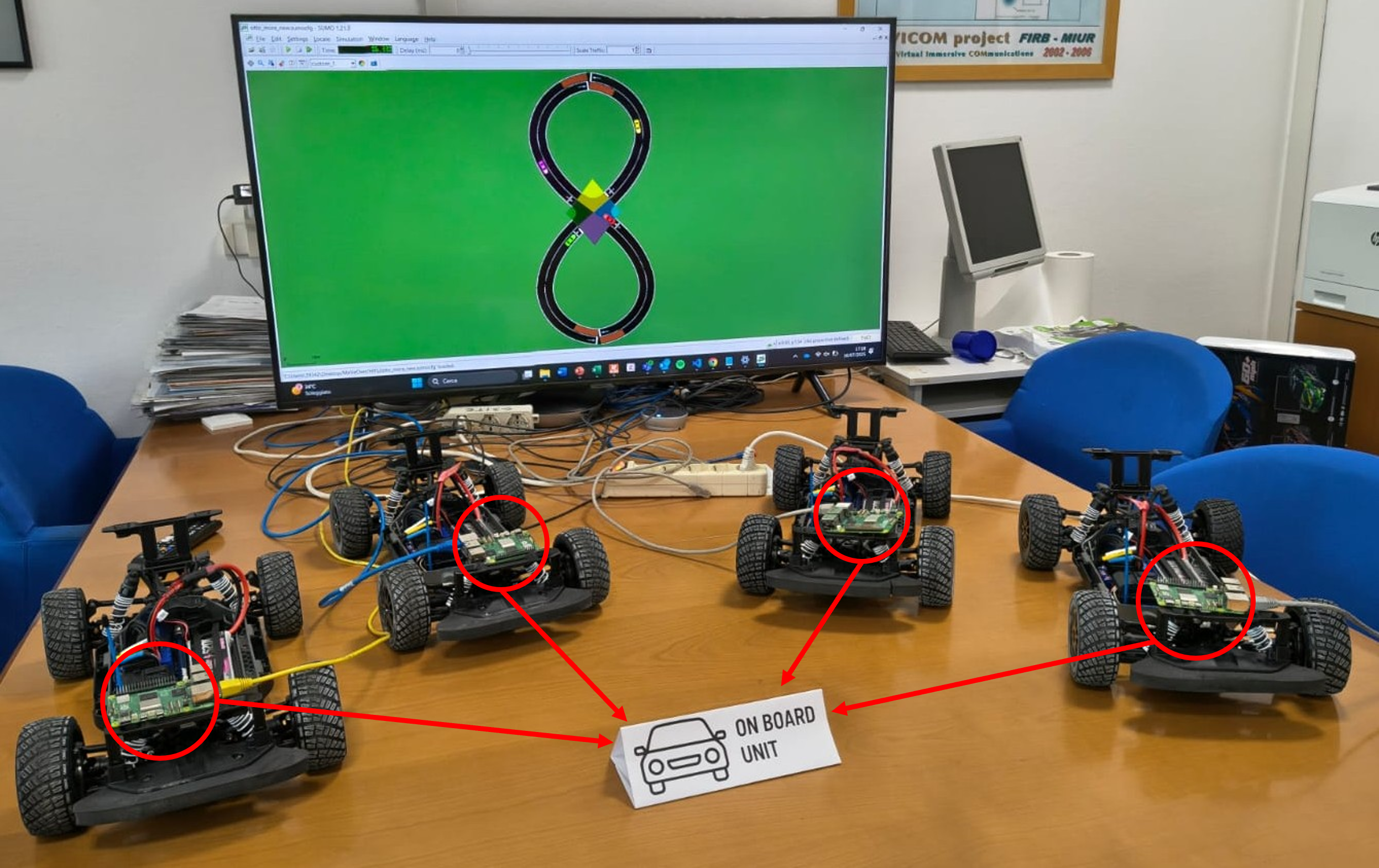}
\label{fig:SUMO}}\\
\subfloat[]{\includegraphics[width=\columnwidth,draft=false]{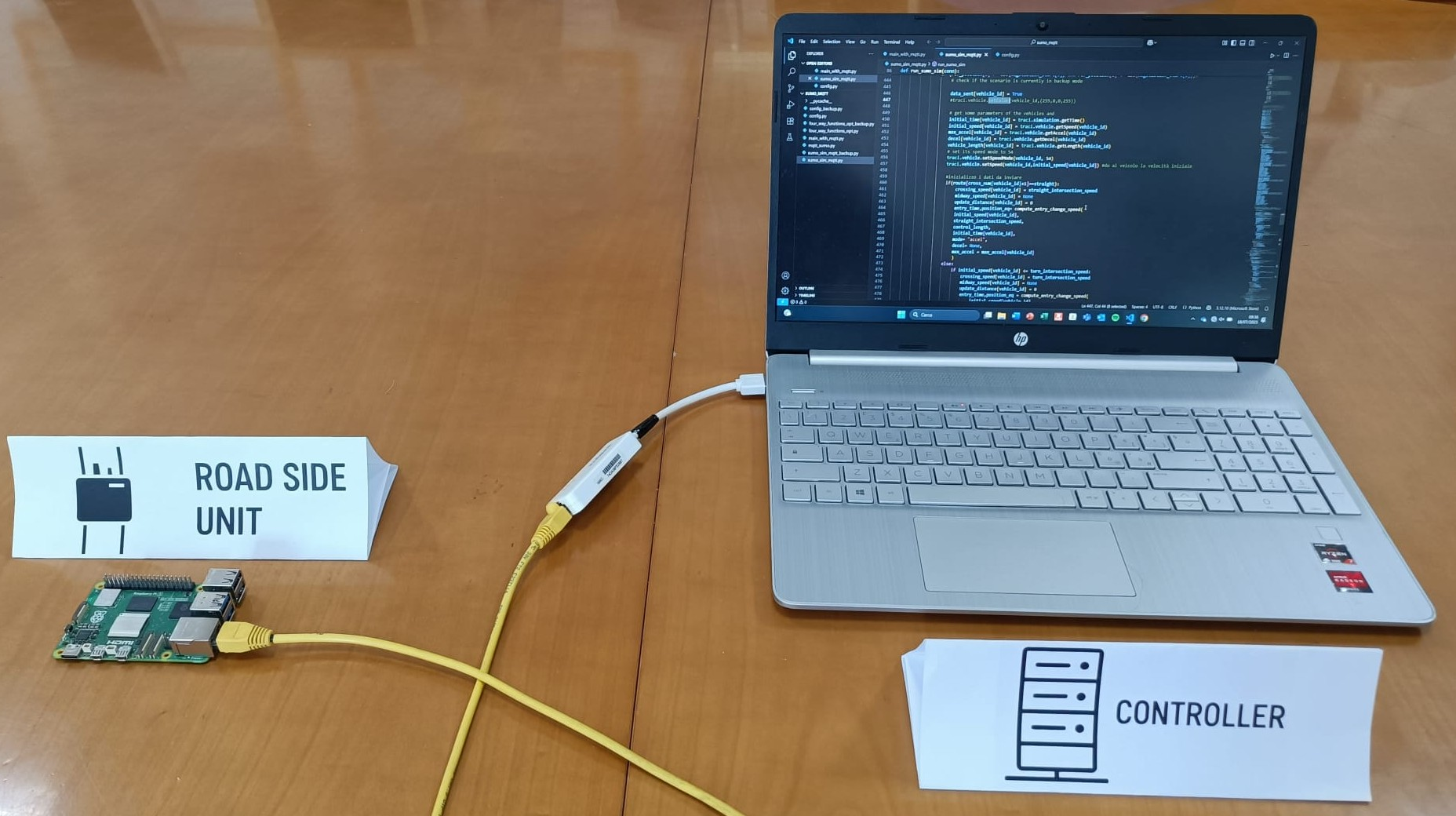}
\label{fig:RSU}}
\caption{PoC. (a) SUMO simulation and OBUs. (b) RSU and controller.}
\label{fig:PoC}
\end{figure}

The demo consists of a number of vehicles moving in a certain scenario, with wireless connectivity with a remote controller. An example of the setup is shown in Fig.~\ref{fig:SUMO} (scenario and OBUs) and Fig.~\ref{fig:RSU} (RSU and controller). In the example shown, the scenario used is an eight-shaped circuit where four vehicles reach the intersection and randomly select the direction to follow, without ever stopping.

Our HiL setup is built upon off-the-shelf and easily customizable hardware and software solutions. The software framework we leverage is a fully open source implementation of the ETSI Cooperative Intelligent Transportation System (C-ITS) network stack, encompassing all the layers from Transport and Networking to Facilities, where different standardized message formats are defined.
This framework, named OScar \cite{OScar_paper_2024}, is written in C++ with a modular structure, enabling a smoother implementation of new features. It is designed to be executed on any Linux-based system, from more complex on-board computers, to low-power embedded systems, including the Raspberry Pi platform we use for our demo \cite{farina2025lowcostcitsstations}. Importantly, OScar supports the provision of positioning data from different sources, i.e., \textit{(i)} the Linux \texttt{gpsd} daemon, that can provide data coming from either virtual or real GNSS receivers, \textit{(ii)} the serial communication port of any (virtual or real) GNSS receiver compatible with the NMEA 0183 standard (supporting also the provision of IMU data via dedicated U-blox messages), \textit{(iii)} JSON traces containing timestamped navigation data that was previously collected in the field. 
In this context, both the solution using the gpsd library, which facilitates the parsing of the data transmitted by gps, and via direct serial communication with gps are optimal, particularly the latter which directly picks up the NMEA sentences with relevant information on vehicle position, speed, and direction.
% \hl{@UNIMORE: do you think it may be relevant to specify here how positioning data can be potentially provided to OScar considering the demo setup, starting from SUMO?}

On the hardware side, we employed a Raspberry Pi 5 Model B, equipped with an ARM Cortex-A76 processor, 8 GB of LPDDR4 RAM and 128 GB of flash storage. The device is powered through a 5 V / 5 A DC USB-C supply and offers a PCIe 2.0 interface, enabling high-speed peripheral expansion. This flexibility allows integration of a wide range of standardized \ac{HAT} modules, such as M.2 SSDs for enhanced storage performance. In our configuration, we installed a PCIe HAT to host a Wi-Fi card tailored to our requirements. Specifically, we selected the Mikrotik R11E-5HND, an IEEE 802.11a/n adapter based on the Qualcomm Atheros AR9580 chipset, featuring an integrated heat sink for improved thermal management. This chipset operates in the upper 5 GHz band and is supported by the ath9k Linux kernel driver, which can be adapted to enable V2X communication under the IEEE 802.11p standard \cite{grazia2018performance}. It is worth noting that the Raspberry Pi 5 typically draws 25 W or more; thus, care must be taken to ensure a suitably rated power supply, particularly when operating on battery power (e.g., via a power bank).

As mentioned earlier, the communication between the \ac{OBU} and the \ac{RSU} is performed through IEEE 802.11p, a version of Wi-Fi specifically designed for the automotive domain. IEEE 802.11p represents, as part of ITS-G5, the main technology of choice for V2X communications in Europe, being already deployed since a few years on Volkswagen vehicles such as the Golf 8 \cite{platoon-paper-cit-vw-2019}, and on the \acp{RSU} of several motorway operators. At the moment, proprietary IP-based messages are exchanged between the OBU and the RSU in our demo. However, as the standardization activities are progressing, we plan to shift to more standardized ETSI \acp{MCM}, which will represent the main enabler for day 3+ applications involving the coordination of multiple connected vehicles \cite{etsi-mcs-pre-std-study}. 
% \hl{@all: do you think we should already specify here the format of the messages we exchange, or would this be part of the description of the algorithm?}
In the current implementation, boards are used as interfaces to forward information between mini-cars and the controller. This exchange of data is achieved by implementing an IP routing system. The routing configuration is manually set using the \texttt{ip route} command, allowing custom definition of the routing tables to ensure efficient and direct packet transmission between network nodes. All of these involved are on the same IEEE 802.11p network and therefore are able to exchange data, due to the 802.11p standard and OCB mode, immediately at very low latency, which is mandatory in the context of V2X applications. 

The algorithm and the related communication protocol implemented in the demo work as follow. When a vehicle, whose position is simulated through SUMO, enters a specific area called \textit{control zone}, it generates a proposal. This proposal contains a possible position, speed, and acceleration profile the vehicle might follow to reach and cross the intersection, ignoring the presence of other vehicles in the scenario. Once generated, the proposal is immediately transmitted to the OBU, which forwards it in unicast to the RSU. Upon reception, the RSU forwards it to the controller, which performs a validation. The validation consists in checking whether the proposed profile conflicts with those already defined for the previous vehicles. If that is the case, the controller replies with a rejection message containing time constraints that the vehicle has to satisfy to avoid conflicts; otherwise, it sends an acceptance notification. The response goes along the same communication chain back to the vehicle. If a rejection is received, the vehicle generates a new proposal, and the procedure is repeated until an acceptance notification is obtained. When the profile is agreed, this is followed by the vehicle and recorded by the controller. In all the cases, the procedure must be completed before the vehicle reaches a given distance from the intersection (which allows the vehicle to safely and comfortably brake); otherwise, the communication with the controller is suspended and the vehicle enters what we call backup mode, meaning that it will not have an agreed speed profile and will only rely on its sensors to cross the intersection. More details about the algorithm can be found in \cite{10575944}.

The demo shows that the algorithm and communication protocol are effective to let the vehicles optimize their speed to approach the intersection without the need to step.

%\hl{communication protocols (polito?)} ... \hl{OScar (polito?)} ... 
%\hl{algorithm (unibo?)} ...

\section{Future work}

\begin{figure}[t]
\centering
\includegraphics[width=80mm]{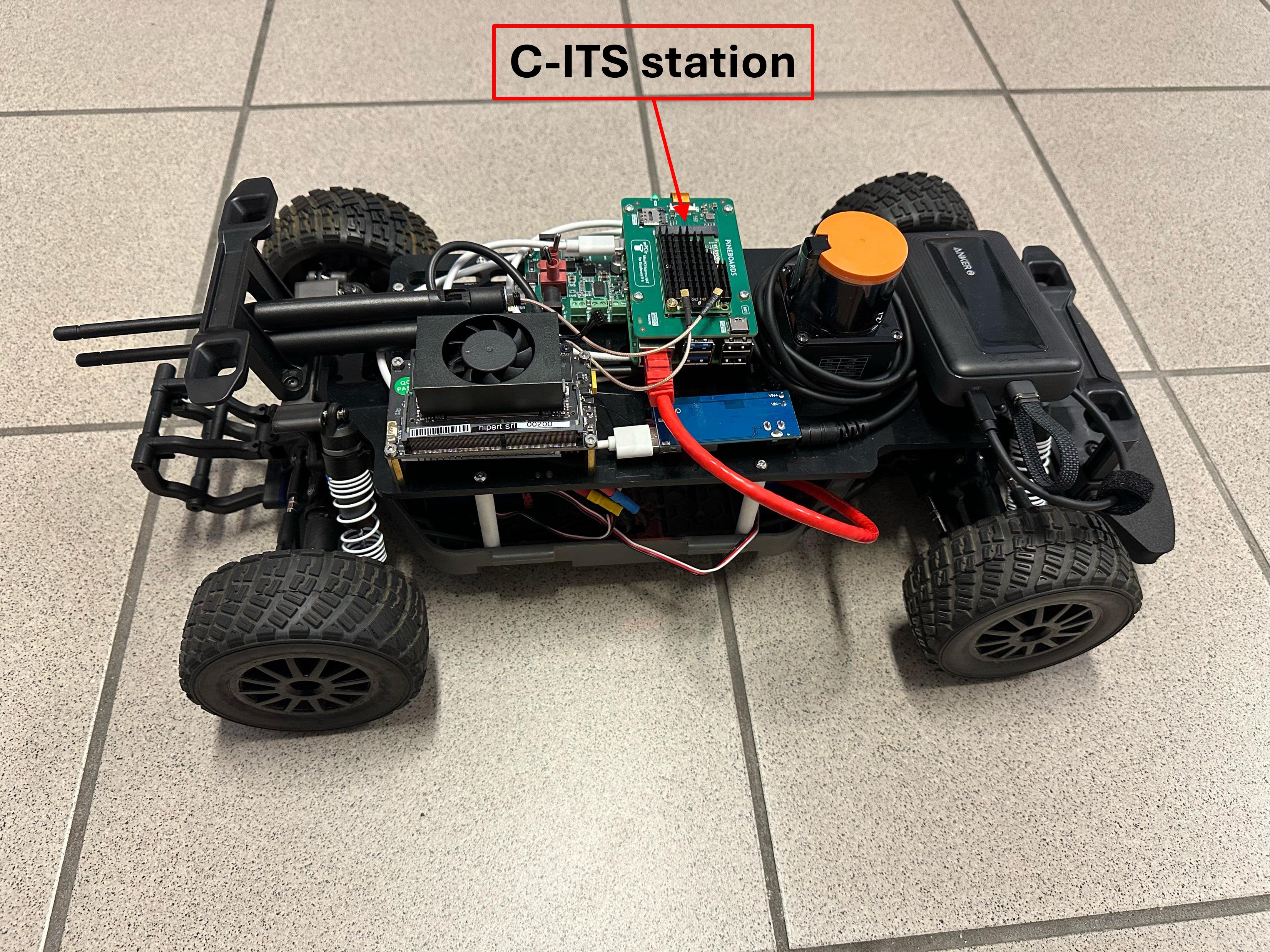}
\caption{Mini-car scale 1:10, equipped with the low-cost OBU.} %\hl{decidiamo se vogliamo evidenziare i box con i componenti, intanto allego la versione raw}}
\label{fig:f1tenth}
\end{figure}

An important aspect to remark is that the use of the Raspberry Pi and OScar, rather then other large and energy-hungry \ac{COTS} \ac{V2X} devices, will allow us to mount the \acp{OBU} over mini-cars 1:10, as shown in Fig.~\ref{fig:f1tenth}. This \ac{HiL} is therefore the preliminary step for a field trial where the mini-cars will move on a track, approaching a real intersection where they will not stop. 

\section*{Acknowledgment} This work was partially supported by the European Union under the Italian National Recovery and Resilience Plan (NRRP) of NextGenerationEU, partnership on ``Telecommunications of the Future'' (PE00000001 - RESTART), project MoVeOver, and national research centre on mobility (CN00000023 - MOST).

%\hl{future works?}

\bibliographystyle{IEEEtran}  
% %\bibliography{biblio,biblio-3GPP}
\bibliography{biblio}

\end{document}

%% file: acronym.tex
\begin{acronym} 
\acro{3GPP}{Third Generation Partnership Project}
\acro{5G}{fifth generation}
\acro{API}{application programming interface}
\acro{C-ITS}{cooperative-intelligent transport systems}
\acro{C-V2X}{cellular-vehicle-to-everything}
\acro{CAM}{cooperative awareness message}
\acro{CAV}{connected and autonomous vehicle}
\acro{COTS}{commercial off-the-shelf}
\acro{CPM}{collective perception message}
\acro{DENM}{decentralized environmental notification message}
\acro{DSRC}{dedicated short range communication}
\acro{ETSI}{European Telecommunications Standards Institute}
\acro{FFT}{fast Fourier transform}
\acro{GNSS}{global navigation satellite system}
\acro{GPS}{global positioning system}
\acro{HAT}{hardware attached-on-top}
\acro{HiL}{hardware-in-the-loop}
\acro{HMI}{human-machine interface}
\acro{IQ}{in-phase and in-quadrature}
\acro{ISAC}{integrated sensing and communication}
\acro{ITS}{intelligent transport system}
\acro{IVIM}{infrastructure to vehicle information message}
\acro{LDM}{local dynamic map}
\acro{LOS}{line-of-sight}
\acro{MAC}{medium access control}
\acro{MCM}{maneuver coordination message}
\acro{MIMO}{multiple input multiple output}
\acro{NLOS}{non-line-of-sight}
\acro{OBU}{on-board unit}
\acro{OFDM}{orthogonal frequency-division multiplexing}
\acro{OScar}{Open Stack for Car}
\acro{PHY}{physical}
\acro{RSU}{road side unit}
\acro{V2N}{vehicle-to-network}
\acro{V2I}{vehicle-to-infrastructure} 
\acro{V2V}{vehicle-to-vehicle} 
\acro{V2X}{vehicle-to-everything} 
\acro{VAM}{vulnerable road user awareness message}
\acro{WAVE}{wireless access in vehicular environment}
\end{acronym}